# A New Technique for Improving Energy Efficiency in 5G Mm-wave Hybrid Precoding Systems

Adeb Salh
*Faculty of Electrical and Electronic Engineering*
*Universiti Tun Hussein Onn Malaysia*
Johor, Malaysia
adebali@uthm.edu.my

Qazwan Abdullah
*Faculty of Electrical and Electronic Engineering*
*Universiti Tun Hussein Onn Malaysia*
Johor, Malaysia
gazwan20062015@gmail.com

Ghasan Hussain
*Department of Electrical Engineering, Faculty of Engineering.University of Kufa, Iraq*
ghasan.alabaichy@uokufa.edu.iq

Razlai Ngah
*Wireless Communication Centre (WCC), Faculty of Engineering, Universiti Technology Malaysia.*
razalingah@utm.my

Lukman Audah
*Faculty of Electrical and Electronic Engineering*
*Universiti Tun Hussein Onn Malaysia*
Johor, Malaysia
hanif@uthm.edu.my

Nor Shahida Mohd Shah
*Faculty of Applied Sciences and Technology, Universiti Tun Hussein Onn Malaysia, Pagoh, Muar, Johor, Malaysia.*
shahida@uthm.edu.my

Shipun Hamzah
*Faculty of Electrical and Electronic Engineering*
*Universiti Tun Hussein Onn Malaysia*
Johor, Malaysia
shipun@uthm.edu.my

*Abstract*— **In this article, we present a new approach to optimizing the energy efficiency of the cost-efficiency of quantized hybrid pre-encoding (HP) design. We present effective alternating minimization algorithms (AMA) based on the zero gradient method to produce completely connected structures (CCSs) and partially connected structures (PCSs). Alternative minimization algorithms offer lower complexity by introducing orthogonal constraints on digital pre-codes to concurrently maximize computing complexity and communication power. As a result, by improving CCS through advanced phase extraction, the alternating minimization technique enhances hybrid pre-encoding. For PCS, the energy-saving ratio grew by 45.3 %, while for CCS, it increased by 18.12 %.**

## I. Introduction

Operations of the massive multi-input-multiple-output (mMIMO) system on millimeter-wave (mm-wave) are aimed at ensuring the expected emergence and growth of traffic as key solutions for fifth generation (5G) high-data multimedia access [1]– [3].

For mm-wave mMIMO systems, the hybrid pre-encoding (HP) architecture is a novel emerging technology that allows the creation of directional beamforming with huge antenna arrays while also increasing energy efficiency (EE). The greater the number of antennas increases the higher the energy cost and hardware complexity of digital pre-code (DPs). Only DP in the baseband can be used to reduce costs, power consumption (PC), and a number of radio frequency (RF) chains [4]. For mMIMO systems, a wide degree of freedom for each RF chain or completely connected structure (CCS) can be activated by phase switches to maximize the user's minimum data [4-6]. The development of four separate algorithms for a single user HP and combiner guarantees the differences between computational complexity and productivity [4]. The author in [5] proposes an iterative update approach for designing a hybrid receiver for phases in RF pre-codes or a combiner. To trade off energy and cost efficiency in RF chains, the CCS [6], [7] develops a heuristic method to solve the problem of baseband precoding and RF chain processing. However, by connecting each RF circuit to a restricted number of antennas, partly connected structures (PCSs) might reduce the complexity of the hardware for practical implementation [6] - [8].

According to [7], increasing the number of chains is unnecessary because it raises the cost and PC. The author offers two hybrid pre-encoding structures for CCS/PCS that improve DP and analogue precoding [7]. Using CCS or PCS with dynamic subarray and low-resolution phase shifters, the communication rate can be increased by adopting the performance and hardware efficiency. By addressing productivity and hardware efficiency, this iterative hybrid beam design can reduce the loss of performance of stationary subarrays coupled to all transmission antennas in each RF circuit [8-10]. However, the CCS and PCS indicators from the above-mentioned gap cannot be achieved compared to the number of data streams $N_s$, because the complexity of the partial relaxation program, the direction-shifting optimization method, and the OMP algorithms are not small, thereby allowing to increase the number of RFs. $N_{RF}$ circuits increase and become inconvenient when evaluating channels for a given RF [11]–[15].

The main objective of this study is to improve the EE by reducing the cost, PC, and quantity of RF based on the proposed efficient alternate minimization algorithms for generating the PC and PC. The CCS of the HP design is performed by updating the phase extraction to constrain zero-gradient based iterative minimization algorithms, provided that $N_{RF}$ is comparable to $N_s$. On the other hand, the PCS-HP matrix is proposed to optimize baseband precoding and reduce hardware complexity by using fewer phase shifters in analog RF precoders, as shown in Fig. 1, to optimize EE and ensure low hardware complexity in mm-wave. In the domain of RF circuits.

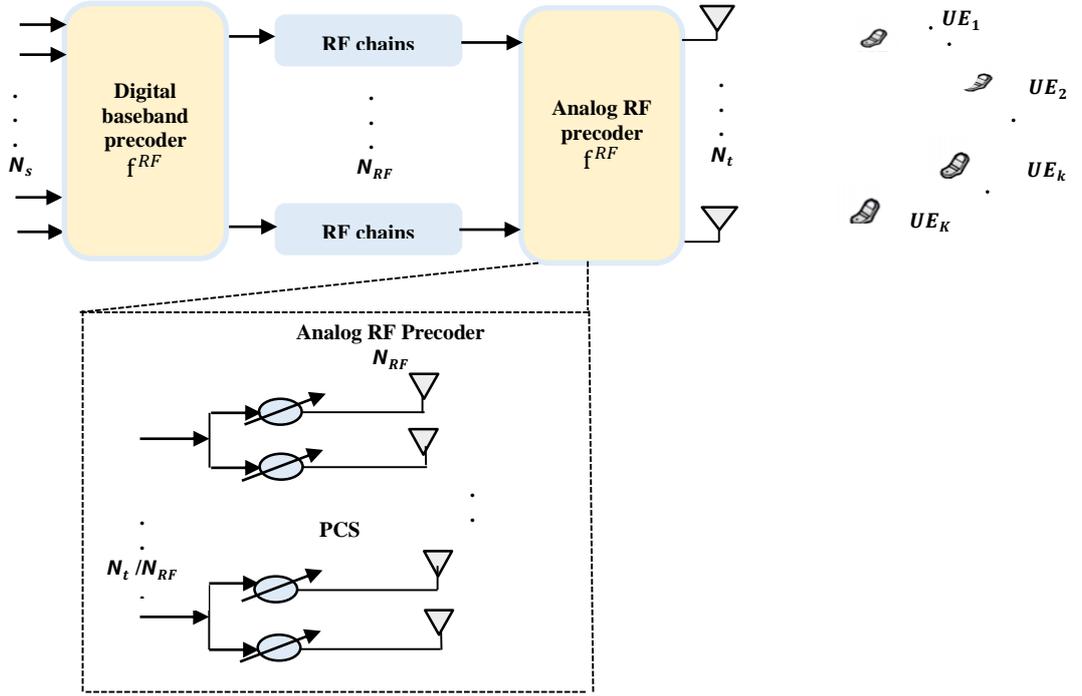

Figure 1. Hybrid precoding in mm-wave MU - mMIMO systems.

## II. SYSTEM MODEL

For the mm-wave multi-user (MU)- mMIMO system of CCS/PCS based on HP, we address the downlink (DL) transmission. The HP is assumed to be the ideal channel state information (CSI) available at both the transmitter and receiver [9] [10]. The data streams $N_s$ send from the transmitter and accumulates $N_t$ transmit antenna and each user equipment (UE) is provided with $N_r$. A single $N_{RF}$ is coupled to an antenna element in each UE. $N_{RF}$ is equipped in the transmitter to support the simultaneous data $N_s$ flow of every antenna element from $N_t$ transmitting antennas to $N_r$ receiving antennas $N_s \leq N_{RF} \ll N_t$. In HP is a crucial task to lower the complexity the received signal at $y_k \in \mathbb{C}^{N_r \times N_t}$ UE can be represented as

$$y_k = g_k^H f^{RF} f_k^{BB} u_k + ó_k \qquad (1)$$

The transmit signal in DL from base station (BS) to user $u_k = [u_1, \dots, u_k \dots, u_K] \in \mathbb{C}^{N_s \times 1}$, and $g_k \in \mathbb{C}^{N_r \times N_t}$ represent the channel matrix corresponding to the $k\,th$ user. $f_k^{BB} \in \mathbb{C}^{N_{RF} \times K}$ is the baseband precoding (Band_BP) matrix can be represent as $f_k^{BB}$. $ó_k \sim \mathcal{CN}(0, \alpha^2)$ is the complex Gaussian noise, and $\alpha_n^2$ represents the noise power. $f^{RF} \in \mathbb{C}^{N_t \times N_{RF}}$. To improve the CSI with L path propagation, show the channel propagation loss associated with that in the low-band channel. The DL channel matrix can be written as follows:

$$g_k = \sqrt{\frac{N_r N_t}{\mathcal{L}_k}} \sum_{l=1}^{\mathcal{L}_k} \Omega_{k,l}\, à(\phi_{k,l}^r)\, à(\phi_{k,l}^t)^T \quad \forall k \qquad (2)$$

where $N_t$ denotes the number of transmit antennas arrayed at the base station, the number of antennas arrayed at the UE is denoted by $N_r$, the UE's complicated channel gain across the lth multi-path is $\Omega_{k,l}$, and the multi-path propagation channels used by the $k\,th$ user is represented by $\mathcal{L}_k$; $l \in [1, 2, \dots \mathcal{L}_k]$. $à_t(\phi_{k,l}^r)$ and $à_t(\phi_{k,l}^t)$. For the azimuth angles of arrival multi-path propagation, represent the normalized transmit and receive array response vectors, respectively. For the N-element of a regular linear array, the array response vector can be expressed as

$$à_{ULA}(\phi_{k,l}^r) = \frac{1}{\sqrt{N}}\left[1, e^{j\frac{2\pi}{\gamma}\sin\phi}, \dots, e^{-j\frac{2\pi d(N-1)}{\gamma}\sin\phi}\right]^T \qquad (3)$$

where $N$ denotes the number of antenna elements with the BS, and $\gamma$ denotes the bandwidth wavelength, and the inter distance between antenna elements is represented by $d = \frac{2}{\gamma}$. The sidelobe beamforming generates an analogue vector based on a signal-to-interference noise ratio. The data rate for UE's $k\,th$ data stream is written as

$$R_k = \varpi \sum_{k=1}^{K} \log_2\left(1 + \frac{\left|g_k^H F_k^{BB}(f_k^{BB})^H f^{RF}(f^{RF})^H g_k\right|^2}{\sum_{\substack{i=1 \\ i \neq k}}^{K} \left|\sum_{k=1}^{K} g_k^H f^{RF} F_i^{BB}(F_i^{BB})^H (f^{RF})^H g_k\right|^2 + \alpha_n^2}\right) \qquad (4)$$

where $\varpi$ is the UE's $k\,th$ UE's bandwidth transmission. The maximum data rate that can be achieved at the $k\,th$ UE as a percentage of the total data rate as $R_{sum} = \sum_{k=1}^{K} R_k$.

## III. PROBLEM FORMULATION

### A. Transmission Power Model

Optimizing Band_BP $f_k^{BB}$ and RF precoding transmission systems in mm-wave MU-mMIMO systems is necessary to maximise the EE at the BS. It is necessary to calculate the RF. The consumption power must be considered in this scenario. Communication power and circuit PC combine up the PC at the transmitter. It can be expressed as

$$p_t = p_{Communi} + p_C \quad (5)$$

In the first term of (5), there are two portions to the communication power $p_{Communi} = p_{PA} + P_{RF}$; $p_{PA}$ the power spent in terms of $P_{RF} = N_{RF} p_{RF}$ represent by the power amplifiers, as well as the power consumed by mixers, filters, and phase shifters at each RF chain. The amount of power used by the power amplifier to convey a signal to the $k\,th$ UE as $p_{PA} = \frac{1}{\Xi}\sum_{k=1}^{K}\|f^{RF} f_k^{BB}\|_F^2$, where $\sum_{k=1}^{K}\|f^{RF} f_k^{BB}\|_F^2$ represent the energy required to transmit signal for $k\,th$ UE, and also represents the efficiency of the power amplifier [11-14]. The complete PC can be written as

$$p_t = \frac{1}{\Xi}\sum_{k=1}^{K}\|f^{RF} f_k^{BB}\|_F^2 + N_{RF} p_{RF} + p_C \quad (6)$$

### B. Energy Efficiency

In the design of the mm-wave MU-mMIMO system, energy efficiency is a major consideration. Maximization of EE depends on the proposed efficient alternate minimization algorithms for generating CCS and PCS. This approach reduces hardware complexity in the RF circuits domain with low-cost phase shifters. The EE maximization problem is an optimization problem. EE

$$EE = \frac{\varpi \sum_{k=1}^{K}\log_2\left(1+\frac{\left|\sum_{k=1}^{K} g_k^H f_k^{BB}(f_k^{BB})^H f^{RF}(f^{RF})^H g_k\right|^2}{\sum_{\substack{i=1 \\ i\neq k}}^{K}\left|\sum_{k=1}^{K} g_k^H f^{RF} f_i^{BB}(f_i^{BB})^H(f^{RF})^H g_k\right|^2 + \alpha_n^2}\right)}{\frac{1}{\Xi}\sum_{k=1}^{K}\|f^{RF} f_k^{BB}\|_F^2 + N_{RF} p_{RF} + p_C} \quad (7)$$

By mutually improving the Band_BP matrix $f_{BB}$ and the RF precoding matrix $f_{RF}$, the energy efficiency may be optimized by attaining a suitable tradeoff between the data rate and the total PC [15],[16]. The foundation of this optimization is

$$\max_{f_{RF}, f_{BB}} EE = \varpi \frac{\sum_{k=1}^{K}\bar{R}_k}{p_t}$$
$$\sum_{k=1}^{K}\|f^{RF} f_k^{BB}\|_F^2 \leq p_{max}$$
$$R_k \geq \Gamma_k, \quad k = 1, \ldots, K \quad (8)$$
$$\|f^{RF}\|_F^2 = \frac{1}{N_t}$$

where $p_{max}$ is the maximum power in the limitations listed above. Constraint (8) states that all UE should be served by a single BS. $\sum_{k=1}^{K}\|f^{RF} f_k^{BB}\|_F^2 \leq p_{max}$ is a non-convex limitation. Because of the matrix variables $f_{RF}$ and $f_{BB}$, the issue in (8), $\|f^{RF}\|_F^2 = 1/N_t$, has trouble addressing antenna selection[17-20].

We suggest that the HP matrix design delivers an optimal solution with less computational cost to tackle this problem. The upper constraint on the EE is derived as shown in the new subsection as.

### C. Upper Bound of EE

Due to the non-concavity of the EE, the EE optimization restrictions are modified in (8) to determine the upper bound of EE. As a result, we apply a zero gradient approach to get the best Band_BP vector. Furthermore, the Band_BP matrix and the RF precoding matrix are used to simplify the derivation, where $f_k^{BB} \in \mathbb{C}^{N_{RF}\times K}$ is connected to the number of RF chains $N_{RF}$ and $f^{RF} \in \mathbb{C}^{N_t \times N_{RF}}$ is connected to the completely DP matrix: $f^{opt} \in \mathbb{C}^{N_t\times K}$, $f^{opt} = f^{RF} f_k^{BB}$ according (6), $f^{opt} = [F_1, F_2, \ldots F_k, \ldots F_K]$, the optimal Band_BP $f^{opt}$ can be written as $f_k$. To find the stationary point of an ideal Band_BP vector for EE, the best solution uses the zero gradient method $EE(F_1, F_2, \ldots F_k)$ and $\bar{R}_k(F_1, F_2, \ldots F_k)$ [21-24].

$$f^{opt} = EE(F_1, F_2, \ldots\ldots F_k) = \arg\max_{f\in \mathbb{C}^{N_t\times K}} EE = \frac{\varpi \bar{R}_k}{\Xi \sum_{k=1}^{K}\|F_j\|^2 + N_{RF} p_{RF} + p_C} \quad (9)$$

The numerator's data rate and the denominator's transmission PC are presented in (10) and (11), respectively, based on (9).

$$\bar{R}_k = \log_2\left(1 + \frac{\left|\sum_{k=1}^{K} g_k^H F_k^{BB}(F_k^{BB})^H g_k\right|^2}{\sum_{\substack{i=1 \\ i\neq k}}^{K}\left|\sum_{k=1}^{K} g_k^H F(F_i^{BB})^H g_k\right|^2 + \alpha_n^2}\right) \quad (10)$$

$$\bar{\mathcal{P}} = \Xi \sum_{k=1}^{K}\|F_k\|_F^2 + N_{RF} p_{RF} + p_C \quad (11)$$

The baseband beamforming is derived from first complete derivation of $EE(F_k) = (\sum_{k=1}^{K}\bar{R}_k/\mathcal{P})$ in (9), the baseband BP $f^{opt} = [F_1, F_2, \ldots, F_k, \ldots F_K]$. The data rate $\bar{R}_k(F_k)$ of the baseband beamforming $f_k$ is calculated using the gradient of $EE(F_k)$.

$$\frac{\partial EE(F_k)}{\partial(F_k)} = \frac{2}{\bar{\mathcal{P}}^2}[\mathbf{Q_k} - \mathcal{T_k}]F_k \quad (12)$$

with

$$\mathbf{Q_k} = \bar{\mathcal{P}}\frac{g_k g_k^H}{\sum_{j=1}^{K} g_k^H F_j(F_j)^H g_k + \alpha_n^2} \quad (13)$$

$$\mathcal{T_k} = \frac{2}{\Xi} I_{N_{RF}}\left(\sum_{i=1}^{K}\bar{R}_i\right) + \frac{2}{\ln 2}\varpi \bar{\mathcal{P}} \sum_{\substack{i=1 \\ i\neq k}}^{K}\frac{g_i^H F_i(F_i)^H g_i}{(\Gamma_i)^2 + \Gamma_i g_i^H F_i(F_i)^H g_i} \cdot g_i g_i^H \quad (14)$$

When the approximate derivation of $\frac{\partial EE(F_k)}{\partial(F_k)} = 0$, the zero-gradient requirement $f_k$ must be satisfied, then the user's local efficiency is stated as

$$\frac{\partial EE(F_k)}{\partial(F_k)} = \frac{2([\mathbf{Q_k} - \mathcal{T_k}]F_k)}{\bar{\mathcal{P}}^2} = 0; \quad 1 \leq k \leq K \quad (15)$$

As illustrated in (8), obtaining the optimized solution is difficult. $EE(F_k)$ has a convergent upper bound, and the upper bound of EE has been achieved [25]. Then we describe the HP challenge and use the CCS/PCS to solve it.

## D. AMA for the CCS Phase Extraction

Using a unitary matrix $\sigma f_{k(:,j)}^{SS}$, the DP matrix should be mutually orthogonal, the same size as $\_f_k^{BB}$ as $\sigma f_{k(:,j)}^{SS}$, as demonstrated in $(f_k^{BB})_{(j,:)}^H f_{k(:,i)}^{BB} = \sigma (f_k^{SS})_{(:,i)}^H \sigma f_{k(:,j)}^{SS}$ (8). In HP, a DP matrix's mutual orthogonal characteristic $f_k^{BB}$ is substituted by $\sigma f_k^{SS}$ (8)

$$(f_k^{SS})_{(:,i)}^H f_{k(:,j)}^{SS} = \begin{cases} \sigma^2 I_{N_s}, & \text{user } k \text{ is servied by BS } j, \ i = j \\ 0, & \text{otherwise } i = 1,2,..I, \ j = 0,1,..J, \ i \neq j \end{cases} \quad (16)$$

$$i, j = 1, 2 \dots N_s$$

In (8) and (16), the objective function is maximized by reducing the ED among $f^{RF} f_k^{BB}$ and $f^{opt}$.

$$\|F^{opt} - F^{RF} F_k^{BB}\|_F^2$$
$$= Tr(F^{opt}(F^{opt})^H)$$
$$- Tr((F^{opt})^H F^{RF} F_k^{BB})$$
$$- Tr((F^{RF})^H F^{op}(F_k^{BB})^H)$$
$$+ Tr((F^{RF})^H F^{RF}(F_k^{BB})^H F_k^{BB})$$
$$\|f^{opt} - f^{RF} f_k^{BB}\|_F^2 = \|f^{op}\|_F^2 - 2\sigma\mathbb{R} Tr(f_k^{SS}(f^{opt})^H f^{RF}) + \sigma^2 \|f^{RF} f_k^{SS}\|_F^2 \quad (17)$$

When the optimization problem $\sigma = \frac{\mathbb{R} Tr(f_k^{SS}(f^{opt})^H f^{RF})}{\|f^{RF} f_k^{SS}\|_F^2}$ is used, the minimum value achieved is (17), the best totally DPs are $o\|f^{op}\|_F^2 - \frac{\{\mathbb{R} Tr(f_k^{SS}(f^{opt})^H f^{RF})\}^2}{\|f^{RF} f_k^{SS}\|_F^2}$. If RF is fixed for the DPs, the phase of $f^{RF}$ is calculated from the phases of a matching precoder selected $f^{opt} f_k^{SS}$. There is only one optimization variable, $f_k^{SS}$, which is equal to

$$\max_{f_k^{SS} \in \mathbb{C}^{N_{RF} \times N_s}} \mathbb{R} Tr(f_k^{SS}(f^{opt})^H f^{RF})$$
$$s.t. \ (f_k^{SS})_{(:,i)}^H f_{k(:,j)}^{SS} = I_{N_s} \quad (18)$$

When $f_k^{SS} = V_1 y^H$, where $((f^{opt})^H f^{RF}) = V \sum y^H$ is CCS in HP, the equality is established merely according to (18). Furthermore, the best HP $(f^{opt})^H f^{RF} = V \mathbb{s} y_1^H$, which is really the SVD of $(f^{opt})^H f^{RF}$. The entries of the diagonal matrix RF and the initial $N_s$ non-zero singular values [27-30]. By decreasing the complication of the phase extraction for AMA, the accurate normalization $f_k^{SS}$ satisfies the power limitation.

## E. HP-AMA for the PCS with a Low-Complexity

According to [4], [9] and [31], every RF in PCS is connected to a certain number of antennas by utilizing a reduced ED between $f^{RF} f_k^{BB}$ and opt, which are utilised to solve and optimise baseband and RF precoding matrices $f_{opt}^{RF}$ and $f_{opt}^{BB}$.

$$\max_{f_{RF} \in \mathbb{C}^{N_t \times f_{RF}}, \ f_{BB} \in \mathbb{C}^{N_{RF} \times K}} \|f^{opt} - f^{RF} f_k^{BB}\|_F^2$$
$$s.t. \ f^{RF} = diag\{F_1, \dots, F_{N_{RF}}\} \quad (19)$$
$$\|f^{RF} f_k^{BB}\|_F^2 \leq p_{max}$$

where $f^{opt}$ denotes the best Band_BP for the $k$ th user.

Every RF chain is equipped with an antenna subarray in accordance with (19) to ensure the overall PC of active RF chains and the size of PCS $f^{RF}$ as

$$f^{RF} = \begin{bmatrix} F_1^{RF} & 0 & 0 \\ 0 & \ddots & 0 \\ 0 & 0 & F_{N_{RF}}^{RF} \end{bmatrix} \quad (20)$$

where $F_i = [i = 1, 2, \dots,] \in \mathbb{C}^{N_{array} \times 1}$ is the analogue precoding vector $f^{RF}$ to the $i$ th subarray, which corresponds to the precoding matrix between both the $i$ th RF chain and the PCS $N_t / N_{RF}$ antennas. We present the AMA for transmission power to optimize baseband due to the varying patterns of the constraint on $f^{RF}$ in the product $f^{RF} f_k^{BB}$ by fixing the $f^{RF}$ as [15], [16], and [32] to produce a satisfactory solution for $f_k^{BB}$.

$$\min_{f_{BB} \in \mathbb{C}^{N_{RF} \times K}} \|f^{opt} - f^{RF} f_k^{BB}\|_F^2 \quad (21)$$
$$s.t. \ \|f^{RF} f_k^{BB}\|_F^2 \leq p_{max}$$

We propose (21) to be a non-convex constraint based on channel in BS array antenna in (2) and the QCQP problem from [15], [17]. Allow $Đ = vec(f_k^{BB})$ and $f^{opt} = vec(f^{opt})$ and $\mathfrak{L} = I_{N_s} \otimes f^{RF}$. The optimal solution $\|f^{opt} - f^{RF} f_k^{BB}\|_F^2 = \|vec(f^{opt}) - f^{RF} f_k^{BB}\|_2^2 = \|vec(f^{opt}) - (I_{N_s} \otimes f^{RF}) f_k^{BB}\|_2^2$. The order restriction to transfer (21) can be expressed in typical QCQP form, which is a challenge in (21).

$$\min_{Đ} \|t f^{opt} - \mathfrak{L} Đ\|_2^2$$
$$s.t. \begin{cases} \|Đ\|_2^2 \leq \frac{p_{max}}{N_t} N_{RF} \\ t^2 = 1 \end{cases} \quad (22)$$

The optimization problem in (22), for example, can be written as

$$\|t f^{opt} - \mathfrak{L} Đ\|_2^2 =$$

$$\underbrace{[Đ^H \ t]}_{T} \underbrace{\begin{bmatrix} (I_{N_s} \otimes f^{RF})^H (I_{N_s} \otimes f^{RF}) & -(I_{N_s} \otimes f^{RF}) f^{opt} \\ -(F^{opt})^H (I_{N_s} \otimes f^{RF}) & (F^{opt})^H f^{opt} \end{bmatrix}}_{\epsilon} \underbrace{\begin{bmatrix} Đ \\ t \end{bmatrix}}_{T} \quad (23)$$

$$\|Đ\|_2^2 = [Đ^H \ t] \underbrace{\begin{bmatrix} I \ p_{max} N_{RF} & 0 \\ 0 & 0 \end{bmatrix}}_{A_1} \underbrace{\begin{bmatrix} Đ \\ t \end{bmatrix}}_{T} \leq \frac{p_{max}}{N_t} N_{RF} \quad (24)$$

$$t^2 = [Đ^H \ t] \underbrace{\begin{bmatrix} 0 \ p_{max} N_{RF} & 0 \\ 0 & 1 \end{bmatrix}}_{A_2} \underbrace{\begin{bmatrix} Đ \\ t \end{bmatrix}}_{T} = 1 \quad (25)$$

The goal function in real QCQP develops from (23). In (22), the feature of HP $\frac{N_{RF}}{N_t}$ allows it to discover the best $N_{RF}$ vectors or the global best solution of the hybrid precoding design problem (24). $\|I_n\|^2$, and $I = T^H T$ are represented by the values of $T = \begin{bmatrix} Đ \\ t \end{bmatrix}$. The original issue may be rewritten as $\|I_n\|^2$ and each difference restriction as

$$\min_{I \in \mathbb{m}^n} T^H \ \epsilon \ T$$

$$s.t.\ \|I_n\|^2 = 1 \quad (26)$$
$$T^H \in T \le \frac{p_{max}}{N_t} N_{RF}$$
$$\text{rank}(I) = 1$$
$$I \ge 0$$

From $T^H \in T = \text{Tr}(T \in T^H)$ is the issue above. The order constraint, which is non-convex with known I, is the most difficult part of (26) to solve. To obtain the flexible version, we first omit (26) for the rank restriction, using limited HP for AMA as

$$\min_{I \in \mathbb{u}^n} \text{Tr}(\in I)$$
$$s.t.\ \text{Tr}(A_1 I) = \frac{p_{max}}{N_t} N_{RF} \quad (27)$$
$$\text{Tr}(A_2 I) = 1$$
$$I \ge 0$$

The set of $n = N_{RF} N_s + 1$ level complex Hermitian matrices is denoted by $\mathbb{u}^n$. Based on preliminary relaxation of the constraint requirement rank(I) = 1, the constraint in (27) remains convex [17], [33]. The decomposition composition between two vectors $I^{opt} = T^H T$, on the other hand, cannot satisfy the optimization solution $I^{opt}$. As a result, the Band_BP $f_k^{BB}$ is addressed by estimating the unconstrained optimal channel estimation of V and $\mathbb{y}^H$ first $N_s$ column[34-38]. For every column of V to represent the principal components the estimated solution is achieved for optimization solution $I^{opt} = \mathbb{E}[T^H T]$. By picking the independent random T that fulfills $\|f^{RF} f_k^{BB}\| \le p_{max}$ by approximate solution, the related solution is achieved (21). The power constraint is

$$\min_{f_{RF} \in \mathbb{C}^{N_{RF} \times N_t}} \|f^{opt} - f^{RF} f_k^{BB}\|_F^2 \quad (28)$$

$$f^{RF} f_k^{BB} = [F_{RF,(:1)}, \ldots, F_{RF,(:N_{RF})}] \begin{bmatrix} F_{k(1,:)}^{BB} \\ \ldots \\ F_{k(N_{RF},:)}^{BB} \end{bmatrix}$$

$$= \sum_i^{N_{RF}} f_{RF,(:N_{RF})} F_{k(N_{RF},:)}^{BB}$$

$$\|f^{opt} - f^{RF} f_k^{BB}\|_F^2 = \frac{N_{RF} \|f_k^{BB}\|_F^2}{N_t} \le p_{max}$$

$$\text{Phase}(f^{opt}) = \text{Phase}(f^{RF} f_k^{BB})$$

To improve the consistency, the value range of the component in the $i\,th$ column of $f^{RF}$ must be continuous (28). By reducing the ED between analogue RF and Band_BP of every component of a matrix, the optimal phase precoders Phase $f^{opt}$ procedure can be obtained with low-cost phase shifters.

## IV. SIMULATION RESULTS

In this section, simulation results offer for establishing EE with multiple RF chains, the multiple antennas, and users in this part. When the number of RF chains for CCS/PCS is increased, EE begins to drop Fig. 2. Even with a great amount of RF chains, the EE using CCS/PCS approaches can achieve the best results. When the number of RF chains is around 4 and 6, the phase extraction of an AMA in CCS offers the optimum EE performance. When the number of RFs is raised, it reaches the close value of the ideal Band_BPs. When the amount of RF chains is between 1 and 6, the proposed approach Band_BP provides greater EE than the two alternative proposed algorithms. The efficiency of EE in relation to the number of transmission antennas $N_t$ is shown in Fig. 3. Based on the computational resources in a mMIMO system, the EE for various precoding techniques lowers as the number of antennas increases. The ideal Band_BP has a higher EE than CCS and PCS when the number of transmission antennas $N_t$ is fixed. In addition to the technical complexity and communication power, the CCS is superior to the PCS in terms of EE.

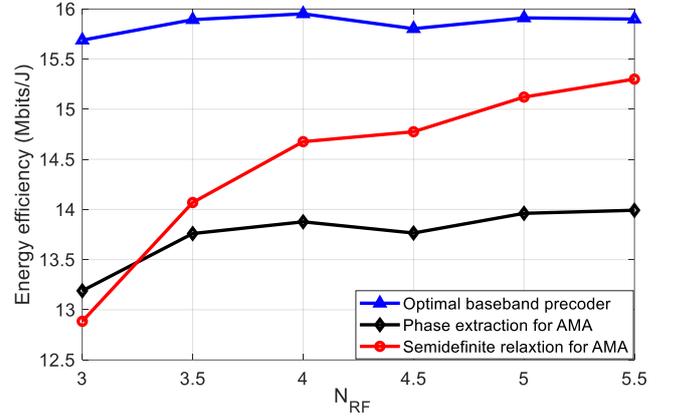

Figure 2. EE in terms of the number of RF chains with various precoding techniques.

The associated power-saving rate for the number of RF chains is depicted in Figure 4. The extraction phase for PC increases when the number of RF chains is considerable CCS rises. In low-complexity HP for AMA, the power-saving ratio (PSR) is 45.3 % when the number of RF chains is 18. The PSR is 18.12% for the CCS of phase extraction as an AMA at the same number of RF chains of 18. When the system attempts to use power gain for interference control, a high PSR is so much more difficult to achieve. When the number of users is small, as indicated in Fig.5, the EE with alternative precoding techniques grows. When the number of RF chains is around 1 and 18, the proposed optimal Band_BPs provide a greater EE than the two different methods. When there are more than six users, the reduced HP in PCS can give enough efficiency and a higher EE over phase recovery in CCS.

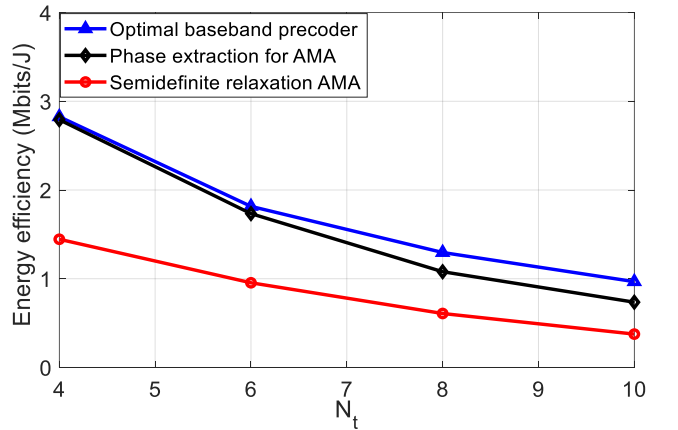

Figure 3. EE vs. $N_t$ number of antennas.

Furthermore, when the number of users is modest, the phase extraction in CCS achieves the close value of the ideal Band_BPs continuously. As a result, the CCS/PCS topologies with variable sub band and reduced power shifters lower the amount of phase shifters while improving the EE. Only when number of transmission antennas is increased, cost efficiency decreases as shown in Fig. 5. When the number of antennas is increased, cost efficiency of ideal Band_BPs increases as well. In the meantime, when the $N_{RF}$ increases, the process recovery in CCS and reduced HP for AMA diminishes. A phase extraction procedure of the matrix $f_k^{opt} (f_k^{SS})_{(:,i)}^H$ is used to attain the highest cost efficiency using CCS structures during every iteration in order to achieve the minimal complexity.

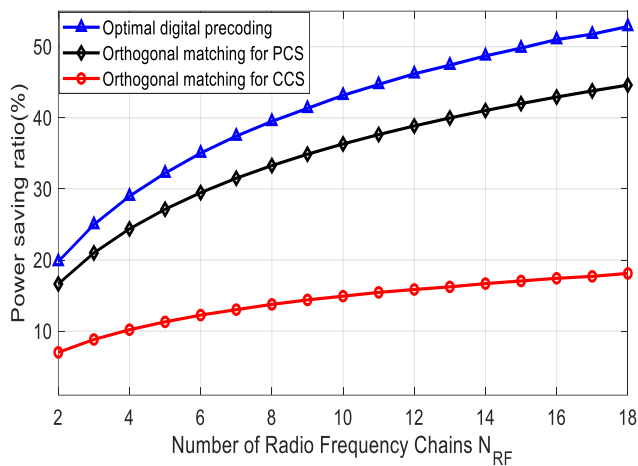

Figure 4. Ratio of power savings to the number of RF chains.

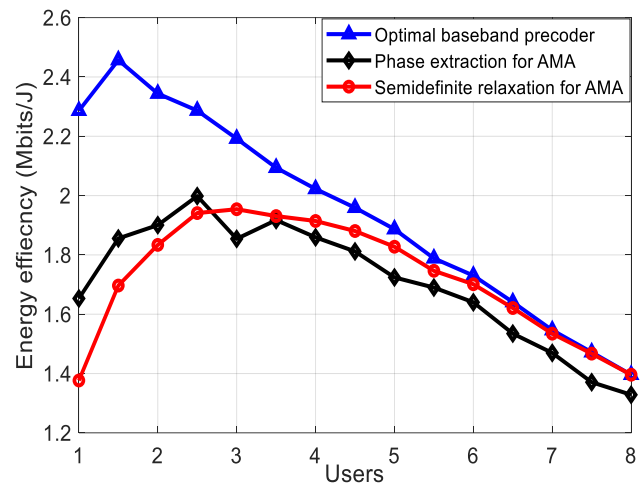

Figure 5. EE's performance in relative to the number of users.

## V. CONCLUSION

In this paper, we present alternating minimization algorithms for an HP-mm-wave mMIMO system to explore the trade-off between energy and cost efficiency. The computing power is taken into account. When the number of RF links is increased, the efficiency of EE degrades. Furthermore, when the complexity of RF chains exceeds the transmitted signal, HPs with CCS can migrate toward the best Band_BP. With a considerable number of RF connections, the PCS enhances EE. The proposed CCS/PCS methods were created to improve the efficiency of the MU-mMIMO communication system. The paper presents a low HP for AMA in CCS/PCS obtains 48.3 % and 17.12 %, respectively, according to simulation findings. In terms of cost efficiency and greatest PSR, the PCS structure outperforms the phase extraction for AMA in CCS.

ACKNOWLEDGMENT

This work was supported by the Ministry of Higher Education Malaysia through the Fundamental Research Grant Scheme FRGS/1/2019/TK04/UTHM/02/8 and Universiti Tun Hussein Onn Malaysia.